\documentclass[aps,prd,superscriptaddress,nofootinbib,amsmath,amsfonts,preprintnumbers,groupedaddress,showpacs,10pt,english]{revtex4}
\usepackage{amsmath}
\usepackage{amssymb}
\usepackage{babel}
\usepackage{wrapfig}
\usepackage{cancel}

\usepackage{relsize,exscale}
\makeatletter

\usepackage{array,multirow,graphicx}
\usepackage{dcolumn}
\usepackage{newlfont}
\usepackage{bm}
\usepackage[colorlinks,citecolor=blue,urlcolor=blue,linkcolor=blue]{hyperref}
\usepackage[figtopcap]{subfigure}
\usepackage{color}
\usepackage{verbatim}

\def\be{\begin{align}}
\def\ee{\end{align}}
\def\bea{\begin{eqnarray}}
\def\eea{\end{eqnarray}}
\def\bal{\begin{align}}
\def\eal{\end{align}}

\usepackage{scalerel}
\usepackage{tikz}
\usetikzlibrary{svg.path}
\definecolor{orcidlogocol}{HTML}{A6CE39}
\tikzset{
 orcidlogo/.pic={
 \fill[orcidlogocol] svg{M256,128c0,70.7-57.3,128-128,128C57.3,256,0,198.7,0,128C0,57.3,57.3,0,128,0C198.7,0,256,57.3,256,128z};
 \fill[white] svg{M86.3,186.2H70.9V79.1h15.4v48.4V186.2z}
 svg{M108.9,79.1h41.6c39.6,0,57,28.3,57,53.6c0,27.5-21.5,53.6-56.8,53.6h-41.8V79.1z M124.3,172.4h24.5c34.9,0,42.9-26.5,42.9-39.7c0-21.5-13.7-39.7-43.7-39.7h-23.7V172.4z}
 svg{M88.7,56.8c0,5.5-4.5,10.1-10.1,10.1c-5.6,0-10.1-4.6-10.1-10.1c0-5.6,4.5-10.1,10.1-10.1C84.2,46.7,88.7,51.3,88.7,56.8z};}}
\newcommand\orcid[1]{\href{https://orcid.org/#1}{\mbox{\scalerel*{
\begin{tikzpicture}[yscale=-1,transform shape]
\pic{orcidlogo};
\end{tikzpicture}
}{|}}}}
\graphicspath{{./}{Figs/}}
\begin{document}
\date{\today}
\title{Special  $N$-dimensional charged anti-de-Sitter black holes in  $f(\mathbb{Q})$ gravitational theory}

\author{G.~G.~L.~Nashed~\orcid{0000-0001-5544-1119}}
\email{nashed@bue.edu.eg}
\affiliation {Centre for Theoretical Physics, The British University, P.O. Box
43, El Sherouk City, Cairo 11837, Egypt}

\begin{abstract}
 In this study, we introduce a toroidal solution for charged anti-de Sitter black holes in $N$ dimensions within the framework of the quadratic form of $f(\mathbb{Q})$ gravity, employing the coincident gauge condition \cite{Heisenberg:2023lru}. We assume $f(\mathbb{Q})$ to take the form $f(\mathbb{Q})=\mathbb{Q}+\frac{1}{2}\alpha \mathbb{Q}^2 - 2\Lambda$, where $N \geq 4$. These black hole solutions are characterized by flat or cylindrical horizons. A notable feature of these solutions is the presence of both electric monopole and quadrupole components in the potential field. These monopole and quadrupole components are inseparable and exhibit interconnected momenta, distinguishing them from the known charged solutions in the linear case of non-metricity theory. Furthermore, we demonstrate that the curvature singularities of these solutions are less severe than those in charged general relativity solutions. Finally, we calculate thermodynamic parameters, including entropy, Hawking temperature, and Gibbs free energy. These calculations confirm the stability of our model.
\end{abstract}

\maketitle
\section{Introduction}\label{S1}

While Einstein's general relativity (GR) has undoubtedly achieved significant success, its limitations in various contexts have become increasingly apparent, prompting scholars to explore alternative theories. In this pursuit, an affine connection compatible with the metric and exhibiting torsion was introduced on flat spacetime, replacing the Levi-Civita connection, which is both torsion-free and metric-compatible and served as the foundation for GR. This modification allowed torsion to play a fundamental role in characterizing gravity. Furthermore, Einstein himself introduced a theory known as metric teleparallel gravity \cite{Unzicker:2005in}. More recently, a new form has emerged in this family: symmetric teleparallel theory. This theory is derived from an affine connection characterized by vanishing curvature and torsion, attributing gravitational effects solely to the non-metricity of spacetime \cite{Nester:1998mp}. In both metric teleparallel theory and its symmetric counterpart, one can construct the torsion scalar $\mathbb{T}$ from the torsion tensor and the non-metricity scalar  $\mathbb{Q}$ respectively. By adopting the Lagrangian $\mathcal{L}=\sqrt{-g}\mathbb{T}$ in the former and $\mathcal{L}=\sqrt{-g}\mathbb{Q}$ in the latter, the corresponding field equations can be derived.   It is important to mention that, despite the successes of  $f(\mathbb{T})$ it encounters issues with local Lorentz transformations and the first law of thermodynamics \cite{Nashed:2021pah, Nashed:2013bfa, Miao:2011ki}. In contrast,  $f(Q)$  does not suffer from these issues. Another notable similarity between $f(\mathbb{T})$ and $f(\mathbb{Q})$ is that neither theory can reproduce a spherically symmetric solution under certain constraints \cite{Nashed:2021pah, Nashed:2018cth, Capozziello:2019uvk, Nashed:2024jqw, DAmbrosio:2021zpm}. Similar to $f(\mathbb{T})$ gravity, $f(\mathbb{Q})$  gravity also leads to deviations from GR. For instance, by setting the function  $\mathbb{Q}$,  we recover the symmetric teleparallel equivalent of GR (STEGR).

The principle of equivalence requires gravity to exhibit a geometric nature. Einstein's GR is a geometric theory of gravity that represents spacetime as a Riemannian manifold. In this framework, the affine connection is the Levi-Civita connection, which is both metric-compatible and torsion-free, and is fully determined by the metric. In GR, the scalar curvature $\mathbb{R}$ is the fundamental quantity that characterizes the manifold. However, the requirement for Riemannian geometry is arbitrary, as a manifold is generally defined by three key geometric attributes: torsion $\mathbb{T}$, curvature $\mathbb{R}$, and non-metricity $\mathbb{Q}$ \cite{BeltranJimenez:2019esp}. Consequently, many gravitational theories can be developed by exploring the relationships between these characteristics.

In metric-affine geometry, specific subcategories emerge based on the presence or absence of certain geometric properties: Riemann-Cartan ($\mathbb{Q}=0$), teleparallel ($\mathbb{R}=0$), and no torsion ($\mathbb{T}=0$). Additional subsets arise when two geometric quantities vanish simultaneously: Weitzenb\o"ck or teleparallel geometry ($\mathbb{R}=\mathbb{Q}=0$), Riemannian geometry ($\mathbb{T}=\mathbb{Q}=0$), and symmetric teleparallel geometry ($\mathbb{R}=\mathbb{T}=0$). When all three quantities vanish, the manifold reduces to the trivial case of a Minkowskian manifold. It is important to note that alternative formulations of GR exist, which are either equivalent to or modifications of GR. One such formulation is the teleparallel equivalent of general relativity (TEGR) \cite{Aldrovandi:2013wha, Nashed:2001im, Maluf:2013gaa, Dialektopoulos:2019mtr, Barros:2020bgg, BeltranJimenez:2019tme, Bajardi:2020fxh, Nashed:2001cp, Ayuso:2020dcu, Flathmann:2020zyj,Nashed:2022zyi, Khyllep:2021pcu, DAmbrosio:2020nev}, characterized by the absence of curvature and non-metricity. Another equivalent formulation is the symmetric teleparallel equivalent of GR (STEGR) \cite{Nester:1998mp, Adak:2004uh, Adak:2005cd, Adak:2008gd, Mol:2014ooa, BeltranJimenez:2017tkd, BeltranJimenez:2018vdo, Gakis:2019rdd}, where both torsion and curvature vanish. In these analogous theories, the Lagrangian density corresponds to their respective scalars, such as $\mathbb{Q}$. Comprehensive analysis and comparisons of these formulations can be found in \cite{Jarv:2018bgs, Capozziello:2022zzh, Heisenberg:2018vsk, Heisenberg:2023lru}.

 Gravitational interactions within the solar system and large-scale cosmic structures have been successfully explained to some extent, yet several unresolved issues persist. These include mysteries such as the unidentified components of the universe (e.g., dark matter and dark energy), challenges related to early-time inflation, and the quantization of gravity. It is now widely recognized that GR (or its equivalent formulations) may not provide the ultimate explanation of gravity, and revisions may be necessary. Consequently, the simplest and most direct modification to address the issues of GR, the teleparallel equivalent of GR (TEGR), or the symmetric teleparallel equivalent of GR (STEGR), is to alter the Lagrangian density, making it a function of the relevant scalars of non-metricity, curvature, or torsion. These approaches have led to the development of theories such as $f(\mathbb{R})$-gravity \cite{DeFelice:2010aj, Sotiriou:2008rp, Nashed:2020kdb, Nashed:2021lzq}, $f(\mathbb{T})$-gravity \cite{Cai:2015emx, Bahamonde:2021gfp, Bamba:2012cp, Nashed:2018piz, Nashed:2018qag, Bamba:2013jqa, Casalino:2020kdr, Nashed:2021pah}, and $f(\mathbb{Q})$-gravity \cite{BeltranJimenez:2017tkd, BeltranJimenez:2018vdo, Nashed:2024pjc, Zhao:2021zab, Lazkoz:2019sjl, Nashed:2024ush, Nashed:2024jmw, Mandal:2020lyq, Capozziello:2022tvv, Capozziello:2022wgl}. Recent analysis have explored the similarities and differences among $f(\mathbb{R})$, $f(\mathbb{T})$, and $f(\mathbb{Q})$ gravity, particularly in terms of degrees of freedom and symmetry breaking \cite{Hu:2022anq}.

Deriving analytical solutions in theories involving high curvature \cite{Sebastiani:2016ras, Vagnozzi:2022moj, Allahyari:2019jqz, Vagnozzi:2019apd} or torsion is often a challenging task, even within the framework of $f(\mathbb{T})$ gravity \cite{Nashed:2013bfa, Capozziello:2012zj, Nashed:2013owg}. Solutions incorporating a cosmological constant exhibit intriguing characteristics, such as the emergence of diverse horizon topologies, in contrast to the asymptotically flat case. In these scenarios, black hole horizons can take spherical, hyperbolic, or planar forms, potentially leading to toroidal or cylindrical shapes depending on the chosen global identifications \cite{Mann:1997iz}. The study of charged black holes in de Sitter (dS) and Anti-de Sitter (AdS) spaces, as well as rotating black holes in four or higher dimensions, has been extensively explored within the frameworks of AdS/CFT and dS/CFT correspondences. For further insights into these investigations, see \cite{Lemos:1994xp, Awad:1999xx, Awad:2000ac, Aharony:1999ti}.

This paper is structured as follows: Section \ref{II} provides a brief overview of the non-metricity formalism and presents the equations of motion for gravity in the context of $f(\mathbb{Q})$ gravity. In Section \ref{III}, we employ a metric potential field with a toroidal configuration in $N$ dimensions to derive the equations of motion for an uncharged black hole solution in $f(\mathbb{Q})$ gravity. The asymptotic behavior of this solution, consistent with Anti-de Sitter (AdS) space, is also discussed in Section \ref{III}. Section \ref{IV} introduces the field equations for charged $f(\mathbb{Q})$ gravity, demonstrating how they yield a precise static black hole solution with charge in AdS space, characterized by both monopole and quadrupole momenta. Section \ref{V} outlines the key physical properties of these charged black holes. In Section \ref{VI}, we delve into the thermodynamics of the black holes, while Section \ref{VII} explores black holes with multiple horizons. Finally, concluding remarks are provided in Section \ref{VIII}.
\section{$f(\mathbb{Q})$-gravitational theory}\label{II}

In this section, we provide a summary of some general features of $f(\mathbb{Q})$ gravitational theory (c.f. \cite{Banerjee:2021mqk,Lin:2021uqa} for for more details).

The expression for the general affine connection on a manifold that is both parallelizable and differentiable is as follows:
\begin{equation}\label{affine}
    \Gamma^\sigma_{\;\mu \nu}= \tilde \Gamma^\sigma_{\;\mu \nu} + K^\sigma_{\;\mu \nu} + L^\sigma_{\;\mu \nu}\,,
\end{equation}
where $\tilde \Gamma^\sigma_{\;\mu \nu}$ is the Levi-Civita connection,  which is characterized by using the metric as\footnote{{Within this study, we adopt relativistic units with c = G = 1. Consequently, the Einstein constant, denoted by $\kappa$, is equivalent to $8\pi$. The symmetric part will be represented by parentheses, denoted as $( )$.  For instance, $A_{(\mu \nu)}$ is defined as $(1/2)(A_{\mu \nu} + A_{\nu \mu})$.  Conversely, the antisymmetric part is denoted by square brackets, represented as [ ], such that $A_{[\mu \nu]}$ is defined as $(1/2)(A_{\mu \nu} - A_{\nu \mu})$.}}:
\begin{equation}\label{Levi-Civita}
  \tilde \Gamma^\sigma_{\;\mu \nu} = \frac{1}{2} g^{\sigma \rho} \left( \partial_\mu g_{\rho \nu} + \partial_\nu g_{\rho \mu}- \partial_\rho g_{\mu \nu}\right)\,.
\end{equation}
Furthermore, $K^\sigma_{\mu \nu}$ represents the contortion which is defined as:
\begin{equation}\label{contortion}
   K^\sigma_{\;\mu \nu}= \frac{1}{2}  T^\sigma_{\;\mu \nu}+ T^{\;\;\;\sigma}_{(\mu \;\; \nu)}\;\;\;\;.
\end{equation}
Here $T^\sigma_{\; \mu \nu}=2 \Gamma^\sigma_{\; [ \mu \nu]}$ is the torsion tensor. Finally,  $L^\sigma_{\;\mu \nu}$ denotes the deformation and is expressed as follows:
\begin{equation}\label{deformation}
   L^\sigma_{\;\mu \nu}= \frac{1}{2} \mathbb{Q}^\sigma_{\;\mu \nu} - \mathbb{Q}^{\;\;\;\sigma}_{(\mu \;\; \nu)}\,.
\end{equation}
Here $ \mathbb{Q}^\sigma_{\;\mu \nu}$  represents  the non-metricity tensor
 expressed as:
\begin{equation}\label{nonmetricity}
   \mathbb{Q}_{\sigma \mu \nu}= \nabla_\sigma g_{\mu \nu}= \partial_\sigma g_{\mu \nu} -\Gamma^\rho_{\;\sigma \mu } g_{\nu \rho} - \Gamma^\rho_{\;\sigma \nu } g_{\mu \rho } \,.
\end{equation}
Consequently, the  scalar of the non-metricity is defined as:
\begin{align}\label{non-m scalar}
   \mathbb{Q}=&g^{\mu \nu}(L^\alpha_{\beta \nu}L^\beta_{\mu \alpha}-L^\beta_{\alpha \beta}L^\alpha_{\mu \nu}) \nonumber\\
     \equiv &-\mathbb{Q}_{\sigma \mu \nu} P^{\sigma \mu \nu} \,.
\end{align}
Here  $P^{\sigma \mu \nu}$ concerning non-metricity conjugate defined as:
\begin{equation}\label{non-m conjugate}
   P^\sigma_{\; \mu \nu} = \frac{1}{4} \left(-\mathbb{Q}^\sigma_{\;\mu \nu}+2Q^{\;\;\;\sigma}_{(\mu \;\; \nu)} + \mathbb{Q}^\sigma g_{\mu \nu}- \tilde{\mathbb{Q}}^\sigma g_{\mu \nu} - \delta^\sigma_{(\mu} \mathbb{Q}_{\nu)}\right)\,.
\end{equation}
Here $\mathbb{Q}_\sigma$ and $\tilde{\mathbb{Q}}_\sigma$ are defined as: \[\mathbb{Q}_\sigma=\mathbb{Q}^{\;\;\mu}_{\sigma\; {\mathbf \cdot} \; \mu}\, \qquad \mbox{and} \quad  \tilde{\mathbb{Q}}_\sigma=\mathbb{Q}^{\mu}_{ \;{\mathbf \cdot}\; \sigma \mu}\,.\]

In the event that both the torsion and the non-metricity equal zero, the connection will resemble the Levi-Civita connection. Torsion and curvature vanish in STEGR gravity, and non-metricity depends on metric and connection.

Modified STEGR was discussed in Ref. \cite{BeltranJimenez:2017tkd} where the action coupled with Maxwell field  yields the form:
\begin{equation}\label{action}
    I=-\frac{1}{2\kappa^2}\int_\mathcal{M}   f(\mathbb{Q}) \sqrt{-g} d^Nx +\int \sqrt{-g} {\cal L}_{ em}~d^{N}x \,.
\end{equation}
Here, $g$, $g_{\mu\nu}$, and $\mathcal M$ represent the determinant, the covariant metric tensor, and the space-time manifold, respectively. The function $f(\mathbb{Q})$ represents the general functional form of the non-metricity scalar $\mathbb{Q}$. The definition of $\kappa$ is given by $\kappa = 2(N-3)\Omega_{N-1} G_N$, where $G_N$ represents the constant of gravitational field in $N$-dimension. In this study,  $\Omega_{N-1}$ denotes the  $(N-1)$-dimensional unit sphere volume. The term of $\Omega_{N-1}$ is specified as   $\Omega_{N-1} = \frac{2\pi^{(N-1)/2}}{\Gamma((N-1)/2)}$. The $\Gamma$-function's dependence is contingent on the dimension of spacetime\footnote{For instance, when $N = 4$, it can be demonstrated that $2(N-3)\Omega_{N-1} = 8 \pi$.}.  {In Eq. (\ref{action}),   ${\cal L}_{
em}=-\frac{1}{2}{ F}\wedge ^{\star}{F}$ represents the Maxwell field Lagrangian,
where  $F = dA$, and  $A=A_{\mu}dx^\mu$, is the 1-form of the  electromagnetic}
potential \cite{Awad:2017tyz}\footnote{ {Charge is fundamentally considered a property of particles that relates to their interactions through fundamental forces, particularly electromagnetism. In classical physics, electric charge is treated as a conserved quantity in isolated systems, typically arising from symmetry principles in the underlying physical laws \cite{Landau:1975pou}.  Noether's theorem is a fundamental result in theoretical physics linking symmetries and conservation laws. It states that every differentiable symmetry of the action of a physical system corresponds to a conservation law. In the case of electric charge, gauge symmetry (specifically U(1) symmetry in quantum field theory) leads to the conservation of charge.  If there is an unknown or hidden symmetry in a physical system, Noether's theorem implies that it might give rise to a new conserved quantity, possibly linked to phenomena not yet fully understood \cite{Peskin:1995ev}. The charge itself can sometimes be a manifestation of a deeper, hidden symmetry not evident at lower energies or scales of the system. This is a typical line of reasoning in advanced physics models like Grand Unified Theories (GUTs) or theories beyond the Standard Model.}}.
When determining the theory's field equations, one conducts separate variations to Eq.~(\ref{action})  with respect to the metric and  matter fields and get  \cite{Heisenberg:2023lru}:
\begin{align}\label{1st EOM}
&\zeta_{\mu \nu}=\frac{2}{\sqrt{-g}} \nabla_\alpha \left( \sqrt{-g} f_\mathbb{Q} P^\alpha_{\;\; \mu \nu }\right) + \frac{1}{2} g_{\mu \nu} f + f_\mathbb{Q} \left( P_{\mu \alpha \beta } \mathbb{Q}^{\;\;\alpha \beta}_\nu
- 2P_{\alpha \beta \mu} \mathbb{Q}^{\;\;\alpha \beta}_\nu\right)+\kappa^2\frac{1}{2}\kappa{{{\cal
T}^{{}^{{}^{^{}{\!\!\!\!\scriptstyle{em}}}}}}}_{\mu \nu}\,,\nonumber\\
&\partial_\nu \left( \sqrt{-g} F^{\mu \nu} \right)=0\;,
\end{align}
{Furthermore, the variation of Eq.~(\ref{action}) with respect to the connection yields:}
\begin{equation}\label{2nd EOM}
{    \nabla^\mu \nabla^\nu \left(\sqrt{-g} f_\mathbb{Q} P^\alpha_{\;\; \mu \nu }\right)=0}\,.
\end{equation}
In this study, ${ {{{\cal T}^{{}^{{}^{^{}{\!\!\!\!\scriptstyle{em}}}}}}}^\nu_\mu}$ is the tensor representing the electromagnetic field's energy-momentum, defined as:
\[
{{{\cal
T}^{{}^{{}^{^{}{\!\!\!\!\scriptstyle{em}}}}}}}^\nu_\mu=F_{\mu \alpha}F^{\nu \alpha}-\frac{1}{4} \delta_\mu{}^\nu F_{\alpha \beta}F^{\alpha \beta}.\]
Here, as is customary, $\mathcal{T}_{\mu \nu}$ denotes the tensor of the energy-momentum of matter, specifically
\begin{equation}\label{EMT}
   \mathcal{T}_{\mu \nu}= -\frac{2}{\sqrt{-g}} \frac{\delta(\sqrt{-g}\mathcal{L}_m)}{\delta g^{\mu \nu}}\;.
\end{equation}
In the given expression, we have $f$ defined as $f(\mathbb{Q})$, and $f_\mathbb{Q}$ as its first derivative concerning $\mathbb{Q}$. It is worth noting that the matter  Lagrangian density is varied independently regarding the connection, resulting in the absence of hyper-momentum. Furthermore, as is widely recognized, the outcomes of GR (in the framework of Symmetric Teleparallel Equivalent of  General Relativity, STEGR) are obtained by setting $f(\mathbb{Q})=\mathbb{Q}$. Consequently, the Lagrangian density takes the form $\mathcal L=-\frac{\mathbb{Q}}{2\kappa^2}+\mathcal L_m$.

\section{Static  AdS/dS black hole solution}\label{III}
We utilize the equations of motions of $f(\mathbb{Q})$ gravity, as expressed by  (\ref{1st EOM}), to investigate the cylindrical  $N$-dimensional spacetime. This analysis  yields the following line element, presented in cylindrical coordinates ($t$, $r$, $\xi_1$, $\xi_2$,$\cdots$ $\xi_{N-2}$), as described in \cite{Awad:2017tyz}:
{\begin{align} \label{le} ds{}^2=h(r)dt^2
-\frac{dr^2}{h_1(r)} -r^2\sum_{i=1}^{N-2}d\xi^2_i\,.\end{align}
Here, $h(r)$ and $h_1(r)$ represent two unknowns dependent on $r$. Furthermore, the equation for $\mathbb{Q}$ concerning the spacetime, as provided by Eq. (\ref{le}), is computed in N-dimensions and yields
\begin{align} \label{Q1}
 \mathbb{Q}=-\frac{(N-2)h_1[(N-3)h+{r}h']}{r^2 h}\,.
\end{align}
Utilizing Eq. (\ref{le}) to the  equations of motions (\ref{1st EOM}) when ${{{\cal
T}^{{}^{{}^{^{}{\!\!\!\!\scriptstyle{em}}}}}}}^\nu_\mu=0$ we obtain the ensuing non-zero components:
\begin{align}\label{df11}
&\zeta_t{}^t\equiv\frac{2(N-2)h_1f_{\mathbb{Q} \mathbb{Q}} \mathbb{Q}'}{r}+\frac{(N-2)f_\mathbb{Q}}{r^2h}\Biggl\{2(N-3)hh_1+rh_1h'+rhh'_1\Biggr\}-f=0\,,\nonumber\\
&\zeta_r{}^r\equiv 2\mathbb{Q}f_\mathbb{Q}-f=0\,,\nonumber\\
&\zeta_{\xi_1}{}^{\xi_1}=\zeta_{\xi_2}{}^{\xi_2}=\cdots \cdots =  \frac{f_{\mathbb{Q}\mathbb{Q}} [r^2\mathbb{Q}+(N-2)(N-3)h_1]\mathbb{Q}'}{(N-2)r}+\frac{f_\mathbb{Q}}{2r^2{h}^2}\Biggl\{2r^2hh_1h''\nonumber\\
&-r^2h_1h'^2+2(2N-5)rhh_1h' +r^2hh'h'_1+2(N-3)h^2[2(N-3)h_1+rh'_1]\Biggr\}-f=0\,.
\end{align}
and the second equation of Eq. (\ref{1st EOM}) takes the form:
\begin{align}
{\frac {2\,rh_1 q'' h+\left(h \left(  h'_1 r+2(N-2)\,h_1 \right)  - h' h_1r \right) q'}{ 2h^{2}r}}=0\,.
\end{align}
Next, we will determine a comprehensive solution to Eqs. (\ref{df11}) by employing a particular shape of $f(\mathbb{Q})$, i.e., \begin{align} \label{fq} f(\mathbb{Q})=\mathbb{Q}+\frac{1}2\alpha \mathbb{Q}^2-2\Lambda\,,\end{align} where $\alpha$ is a dimensional constant  that has the unite of ${\textit length^2}$ and $\Lambda$ is the cosmological constant.
In the context of this particular configuration of $f(\mathbb{Q})$,  Eqs. (\ref{Q1}) yields:
\begin{align}\label{fe}
&\zeta_t{}^t\equiv \frac{2(N-2)\alpha h_1\mathbb{Q}'}{r}+\frac{(1+\alpha \mathbb{Q})(N-2)}{r^2h}\Biggl\{2(N-3)hh_1+rh_1h'+rhh'_1\Biggr\}-\mathbb{Q}-\frac{\alpha}2 \mathbb{Q}^2+2\Lambda=0\,,\nonumber\\
&\zeta_r{}^r\equiv  \mathbb{Q}+\frac{3\alpha}2  \mathbb{Q}^2+2\Lambda=0\,,\nonumber\\
&\zeta_{\xi_1}{}^{\xi_1}\equiv\zeta_{\xi_2}{}^{\xi_2}=\cdots \cdots =\zeta_{\xi_{N-2}}{}^{\xi_{N-2}}=\frac{\alpha[r^2\mathbb{Q}+(N-2)(N-3)h_1]\mathbb{Q}'}{(N-2)r}+\frac{(1+\alpha \mathbb{Q})}{2r^2{h}^2}\Biggl\{2r^2hh_1h''\nonumber\\
& -r^2h_1h'^2+2(2N-5)rhh_1h'+r^2hh'h'_1+2(N-3)h^2[2(N-3)h_1+rh'_1]\Biggr\} -\mathbb{Q}-\frac{\alpha}2 \mathbb{Q}^2+2\Lambda=0\,.
\end{align}
A solution in the general $N$-dimensional case for Eq. (\ref{fe}) is:
\begin{eqnarray}\label{st}
& &  h(r)=-\frac{r^2(1\pm\sqrt{1-12\alpha \Lambda})}{3(N-1)(N-2)\alpha}+\frac{c_1}{r^{N-3}}, \qquad \quad
h_1(r)=h(r)\,.  \nonumber\\
\end{eqnarray}}
{ In this context, $c_1$ represents a dimensional constant of integration which is related to the gravitational mass of the system.} For the sake of simplifying the calculations, we will make the assumption that \begin{align}\label{cons}\Lambda=\frac{1}{12\alpha}\,.\end{align} {  Assumption (\ref{cons}) leads to a special
solution that does not have a general relativity limit since the dimensional constant $\alpha \neq0$. Moreover, this assumption yields a unique solution that can be expressed in the form:}\footnote{These solutions exhibit two possible values for the cosmological constant: $\frac{1 \pm \sqrt{1-12\alpha \Lambda}}{12 \alpha}$.}
\begin{eqnarray}\label{st11}
& &  h(r)=-\frac{r^2}{3(N-1)(N-2)\alpha}+\frac{c_1}{r^{N-3}}\,.
\end{eqnarray}

{ Equation (\ref{st11}) clearly illustrates that the higher order of $f(\mathbb{Q})$ in the case of quadratic form acts as a cosmological constant. Moreover, solution (\ref{st11})  also solve (\ref{2nd EOM}) automatically \cite{Heisenberg:2022mbo}.}
\section{Charged AdS/dS black hole solution}\label{IV}
Utilizing Eq. (\ref{le}) to the  equations of motions (\ref{1st EOM}) as ${{{\cal
T}^{{}^{{}^{^{}{\!\!\!\!\scriptstyle{em}}}}}}}^\nu_\mu\neq 0$ we obtain the subsequent non-zero  components as:{\begin{align}\label{df1}
&\zeta_t{}^t\equiv\ \frac{2(N-2)h_1f_{\mathbb{Q}\mathbb{Q}} \mathbb{Q}'}{r}+\frac{(N-2)f_\mathbb{Q}[2(N-3)hh_1+rh_1h'+rhh'_1]}{r^2h}-f+\frac{2q'^2(r)h_1}{h}=0\,,\nonumber\\
&\zeta_r{}^r\equiv2\mathbb{Q}f_\mathbb{Q}-f+\frac{2q'^2(r)h_1}{h}=0\,,\nonumber\\
&\zeta_{\phi_1}{}^{\phi_1}=\zeta_{\phi_2}{}^{\phi_2}=\cdots \cdots =\zeta_{\phi_{N-2}}{}^{\phi_{N-2}}\equiv \frac{f_{\mathbb{Q}\mathbb{Q}} [r^2\mathbb{Q}+(N-2)(N-3)h_1]\mathbb{Q}'}{(N-2)r}+\frac{f_\mathbb{Q}}{2r^2{h}^2}\Biggl\{2r^2hh_1h'' -r^2h_1h'^2\nonumber\\
&+4(N-3)^2h^2h_1+2(2N-5)rhh_1h'+r^2hh'h'_1+2(N-3)rN^2N'_1\Biggr\}-f-\frac{2q'^2(r)h_1}{h}=0\,.
\end{align}
Finally, Eq. (\ref{1st EOM}) yields:
\begin{align}\label{fc}
\frac{rq'[h_1h'-hh'_1]-2rhh_1q''-2(N-2)hh_1q'}{2rh^2}=0\,.
\end{align}
We will now seek a comprehensive solution to the aforementioned differential equations.? By employing a particular format for $f(\mathbb{Q})$ defined by Eq. (\ref{fq}) we get:
\begin{align}\label{fec}
&\zeta_t{}^t\equiv \frac{2(N-2)\alpha h_1\mathbb{Q}'}{r}+\frac{(1+\alpha \mathbb{Q})(N-2)}{r^2h}\Biggl\{2(N-3)hh_1+rh_1h'+rhh'_1\Biggr\}-\mathbb{Q}-\frac{\alpha}2 \mathbb{Q}^2+2\Lambda+\frac{2q'^2(r)h_1}{h}=0\,,\nonumber\\
&\zeta_r{}^r\equiv  \mathbb{Q}+\frac{3\alpha}2  \mathbb{Q}^2+2\Lambda+\frac{2q'^2(r)h_1}{h}=0\,,\nonumber\\
&\zeta_{\xi_1}{}^{\xi_1}\equiv\zeta_{\xi_2}{}^{\xi_2}=\cdots \cdots =\zeta_{\xi_{N-2}}{}^{\xi_{N-2}}=\frac{\alpha[r^2\mathbb{Q}+(N-2)(N-3)h_1]\mathbb{Q}'}{(N-2)r}+\frac{(1+\alpha \mathbb{Q})}{2r^2{h}^2}\Biggl\{2r^2hh_1h'' -r^2h_1h'^2\nonumber\\
&+2(2N-5)rhh_1h'+r^2hh'h'_1+2(N-3)h^2[2(N-3)h_1+rh'_1]\Biggr\} -\mathbb{Q}-\frac{\alpha}2 \mathbb{Q}^2+2\Lambda-\frac{2q'^2(r)h_1}{h}=0\,.
\end{align}}
{A general $N$-dimension  solution  of Eqs. (\ref{fc}) and (\ref{fec})  has the form:
\begin{align} \label{solc}
&h(r)=\frac{r^2(N-3)^4c_2{}^4}{(N-1)(N-2)(2N-5)^2c_3{}^2}+\frac{c_1}{r^{N-3}}+\frac{3(N-3)c_2{}^2}{(N-2)r^{2(N-3)}}
+\frac{2(N-3)c_2c_3}{(N-2)
r^{3N-8}},\nonumber\\
&
 h_1(r)= h(r)h_2(r), \qquad h_2(r)=-\frac{c_3{}^2(2N-5)^2}{3\alpha(N-3)^4c_2{}^4\left[1+\frac{(2N-5)c_3}{c_2(N-3)r^{N-2}}\right]^2},\qquad q(r)=\frac{c_2}{r^{N-3}}+\frac{c_3}{r^{2N-5}}\,.\nonumber\\
\end{align}
Here, we assign the values of $c_1$ and $c_2$ as $M= -c_1$, $v=c_2$, which signifies the momentum associated with the monopole.  The expression for the quadrupole momentum is given by:\begin{align} \label{const}c_3{}^2= -\frac{3 \alpha (N-3)^4c_2{}^4}{(2N-5)^2}.\end{align} It is worth noting that Eq. (\ref{const}) indicates that $\alpha$ must have a value in the negative range; otherwise, an impractical solution is derived. As a result, the monopole momentum is related to the
quadrupole momentum of the solution. In this case, one gets

\begin{align} \label{solc}
&h(r)=\frac{r^2}{3(N-1)(N-2)\left|\alpha\right|}-\frac{M}{r^{N-3}}+\frac{3(N-3)v^2}{(N-2)r^{2(N-3)}}+\frac{2\sqrt{3\left|\alpha\right|}(N-3)^3v^3}{(2N-5)(N-2)
r^{3N-8}},\nonumber\\
&
 h_1(r)= h(r)h_2(r), \qquad h_2(r)=\frac{1}{\left[1+\frac{v(N-3)\sqrt{3\left|\alpha\right|}}{r^{N-2}}\right]^2},\qquad q(r)=\frac{v}{r^{N-3}}+\frac{(N-3)^2v{}^2\sqrt{3\left|\alpha\right|}}{(2N-5)r^{2N-5}}\,.\nonumber\\
\end{align}
As evident from Eq. (\ref{solc}), the electric potential $q(r)$ is contingent on both monopole and quadrupole moments. When setting $v=0$, both moments disappear, resulting in a non-charged solution. This implies that the quadratic form of $f(\mathbb{Q})$ does not allow for a Reissner-Nordstr\"om solution within the linear framework of the non-metricity theory.}
\section{The key characteristics of the black hole solutions (\ref{solc}) }\label{V}
Now, let's examine some pertinent aspects of the solution with the charge outlined in the preceding section. The line-element of the solution (\ref{solc}) is expressed as:
{\begin{align}\label{metl}
&ds{}^2=\Biggl[\frac{r^2\Lambda_{eff}}{3}-\frac{M}{r^{N-3}}+\frac{3(N-3)v^2}{(N-2)r^{2(N-3)}}+\frac{2\sqrt{3\left|\alpha\right|}(N-3)^3v^3}{(2N-5)(N-2)
r^{3N-8}}
\Biggr]dt^2\nonumber\\
-&\frac{dr^2}{\left[1+\frac{v(N-3)\sqrt{3\left|\alpha\right|}}{r^{N-2}}\right]^2\Biggr[\frac{r^2\Lambda_{eff}}{3}-\frac{M}{r^{N-3}}+\frac{3(N-3)v^2}{(N-2)r^{2(N-3)}}
+\frac{2\sqrt{3\left|\alpha\right|}(N-3)^3v^3}{(2N-5)(N-2)
r^{3N-8}}\Biggr]} -r^2\sum_{i=1}^{N-2}d\xi^2_i\,.\nonumber\\
&\end{align}
Here $\Lambda_{eff}=\frac{1}{(N-1)(N-2)\left|\alpha\right|}$\footnote{{In the case of four-dimensions $\Lambda_{eff}=\frac{1}{6\left|\alpha\right|}$.}}}. Equation (\ref{solc}) clearly indicates that the line element of the charged solution asymptotically approaches AdS spacetime. It is noteworthy that there is no analogous solution in the linear form of non-metricity theory as  $\alpha \rightarrow 0$. This suggests that the charged solution has no counterpart in the lower-order terms of non-metricity theory. In the limit $v \rightarrow 0$, we recover the non-charged black holes with asymptotically AdS behavior, as discussed earlier. It is important to mention that, despite the differences in the metric components, particularly $g_{tt}$ and $g^{rr}$, these asymptotically AdS-charged solutions share common event and Killing horizons.

\underline{Singularity:}\vspace{0.2cm}\\
In this context, we identify tangible singularities by evaluating all relevant invariants within the framework of
  $f(\mathbb{Q})$ theory. Since the function $h(r)$ may possess roots, denoted a $r_n$, it is crucial to analyze the behavior of the invariants near these roots. Upon computing the various invariants, we find the following:
\begin{align}\label{Inv}
& \mathbb{Q}=\frac{1}{3\left|\alpha\right|}+\frac{2(N-3)v}{3\left|\alpha\right| r^{N-2}}\,,\nonumber\\
& \mathbb{R}^{\mu \nu \lambda \rho}\mathbb{R}_{\mu \nu \lambda \rho}\approx\frac{2N}{9(N-1)(N-2)^2\alpha^2}-\frac{16(N-3)v}{9(N-2)^2(N-1)\sqrt{\left|\alpha\right|^3}r^{N-2}}+{\cal O}\left(\frac{1}{r^{2(N-2)}}\right)\,, \nonumber\\
& \mathbb{R}^{\mu \nu}\mathbb{R}_{\mu \nu }\approx\frac{N}{9(N-2)^2\alpha^2}-\frac{4(N-3)v}{3(N-2)^2\sqrt{\left|\alpha\right|^3}r^{N-2}}+{\cal O}\left(\frac{1}{r^{2(N-2)}}\right)\,, \nonumber\\
& \mathbb{R}\approx-\frac{N}{3(N-2)\alpha}+\frac{4(N-3)v}{3(N-2)\sqrt{\left|\alpha\right|^3}r^{N-2}}+{\cal O}\left(\frac{1}{r^{2N-3}}\right)\,,
 \nonumber\\
&\mathbb{Q}^{\mu \nu \rho}\mathbb{Q}_{\mu \nu \rho}\approx \frac{4N}{3(N-1)(N-2)\alpha}-\frac{16(N-3)v}{3(N-2)(N-1)\sqrt{\left|\alpha\right|}r^{N-2}}+{\cal O}\left(\frac{1}{r^{N-1}}\right)\,, \nonumber\\
%
%
&{\mathbb{Q}}_{\mu}{ \mathbb{Q}}^{\mu}\approx \frac{4(N-2)}{3(N-1)\left|\alpha\right|}-\frac{16(N-2)(N-3)v}{3(N-1)\sqrt{\left|\alpha\right|}r^{(N-2)}}+{\cal O}\left(\frac{1}{r^{(N-1)}}\right) \,,\nonumber\\
&\tilde{\mathbb{Q}}_{\mu}\tilde{\mathbb{Q}}^{\mu}\approx \frac{4(N-2)(N-3)^2v^2}{27(N-1)\alpha^2r^{2(N-2)}}+{\cal O}\left(\frac{1}{r^{3(N-2)}}\right)\,,
\end{align}
where $\mathbb{R}^{\mu \nu \lambda \rho}\mathbb{R}_{\mu \nu \lambda \rho}$, $\mathbb{R}^{\mu \nu}\mathbb{\mathbb{Q}}_{\mu \nu }$, $\mathbb{Q}$, $\mathbb{Q}^{\mu \nu \lambda}\mathbb{Q}_{\mu \nu \lambda}$ $\mathbb{Q}^{\mu }\mathbb{Q}_{\mu }$, $\tilde{\mathbb{Q}}^{\mu }\tilde{\mathbb{Q}}_{\mu }$ and $\mathbb{Q}$ are all the possible invariants that can be constructed in this theory\footnote{ Here Kretschmann scalar,  the Ricci tensor square, the Ricci scalar, the non-metricity tensor square,  the non-metricity square vectors and the non-metricity scalar, are defined as:  $\mathbb{R}^{\mu \nu \lambda \rho}\mathbb{R}_{\mu \nu \lambda \rho}$, $\mathbb{R}^{\mu \nu}\mathbb{\mathbb{Q}}_{\mu \nu }$, $\mathbb{Q}$, $\mathbb{Q}^{\mu \nu \lambda}\mathbb{Q}_{\mu \nu \lambda}$ $\mathbb{Q}^{\mu }\mathbb{Q}_{\mu }$, $\tilde{\mathbb{Q}}^{\mu }\tilde{\mathbb{Q}}_{\mu }$ and $\mathbb{Q}$,  respectively}.
 The aforementioned invariants reveal the following:\vspace{0.1cm}\\
 a) A singularity exists at $r=0$, characterized as   curvature singularity. \vspace{0.1cm}\\
 b)In the charged solution, the singularity is indicated by the non-metricity scalar at $r=0$. Near $r=0$, for the charged case, the behavior of the invariants is as follows:   $K=\mathbb{R}_{\mu \nu}\mathbb{R}^{\mu \nu} =\mathbb{R}=\mathbb{Q}^{\alpha \beta \gamma}\mathbb{Q}_{\alpha \beta \gamma}=\mathbb{Q}^{\alpha}\mathbb{Q}_{\alpha}=\tilde{\mathbb{Q}}^{\alpha}\tilde{\mathbb{Q}}_{\alpha}=\mathbb{Q}\sim r^{-(N-2)}$. This is in contrast to the solutions in the non-metricity-Maxwell theory, which exhibit  $K = r^{-3(N-2)}$, and $\mathbb{Q}^{\alpha \beta \gamma}\mathbb{Q}_{\alpha \beta \gamma}=\mathbb{Q}^{\alpha}\mathbb{Q}_{\alpha}\approx r^{-(N-1)}$. This clearly demonstrates that the singularity in the charged case is significantly less severe than that of the linear form of non-metricity for the same solution.This outcome raises an intriguing question: Do these singularities correspond to "weak singularities" as defined by Tipler and Krolak \cite{Tipler:1977zza, clarke1985conditions}? Moreover, it prompts further investigation into whether geodesics can be extended beyond these singularities, a subject to be addressed in future studies.  In conclusion, the invariants particularly those from GR, show that the charged solution differs notably from its GR counterpart.

\section{The thermodynamic characteristics of the black hole solution}\label{VI}
To investigate the thermodynamic characteristics of the newly discovered solution specified in Eq.~(\ref{solc}), we introduced the concept of the Hawking temperature \cite{Nashed:2023qjm,Mazharimousavi:2023tbn}\footnote{{Due to the inequality of the metric potentials in the solution given by Eq. (\ref{solc}), i.e., $h\neq h_1$, the Hawking temperature deviates from the conventional one where the metric potentials are equal  \cite{Sheykhi:2012zz,Sheykhi:2010zz,Hendi:2010gq,Nashed:2023qjm,Sheykhi:2009pf, Nashed:2022mij,Nashed:2021ctg,Nashed:2021lzq,Nashed:2021pah,Nashed:2020kdb}. Equation \eqref{Temp} coincides with the ones presented in literature  when $h\neq h_1$ since $hh_1=-1$.}} as:
  \begin{equation}\label{Temp}
T_2 = \frac{h'(r_2)}{4\pi\sqrt{-h(r_2)h_1(r_2)}}\,.
\end{equation}
In this context, the symbol $'$ indicates a derivative with respect to the event horizon, situated at the value of r equal to $r_2$. This value of $r_2$ represents the most significant positive root of  $h_1(r_2) = 0$, while ensuring that $h'_1(r_2)$ is not equal to zero.
The Bekenstein-Hawking entropy of $f(\mathbb{Q})$  theory is represented as \cite{Cognola:2011nj,Zheng:2018fyn}\footnote{{ We should keep in mind that the entropy framed in linear non-metricity theory differs from that framed in $f(\mathbb{Q})$  theory.  When  $f(\mathbb{Q})=\mathbb{Q}$ Eq. \eqref{ent} will be the standard definition of entropy. However, when  $f(\mathbb{Q})\neq \mathbb{Q}$ we should take into account the effect of the higher order of non-metricity theory similar to the case of $f(R)$ \cite{Nashed:2021lzq,Nashed:2021pah}}}:
\begin{equation}\label{ent}
S(r_2)=\frac{1}{4}A{f_\mathbb{Q}}(r_2)\,,
\end{equation}
{where in this study ${f_\mathbb{Q}}=1+\alpha {\mathbb{Q}}$ as given by Eq. (\ref{Inv}).}
In this frame, $A$ represents the surface area of the event horizon. The black hole will be thermodynamically stable according to the heat capacity indicator, which is denoted as $C_1$; if $C_1>0$, then it is stable, and if $C_1<0$, then it will not be stable. In the subsequent analysis, we evaluate the thermal stability of these black hole solutions by observing the behavior of their heat capacities \cite{Nouicer:2007pu,Chamblin:1999tk}
\begin{equation}\label{m55}
C_2=\frac{dE_2}{dT_2}= \frac{\partial M}{\partial r_2} \left(\frac{\partial T_2}{\partial r_2}\right)^{-1}\,.
\end{equation}
In this context, $E_1$ represents the energy.   Ultimately, the Gibbs free energy is established as follows\cite{Zheng:2018fyn,Kim:2012cma}:
\begin{equation}
\label{enr}
G_2= E_2- T_2S_2 \, .
\end{equation}

\begin{figure*}
\centering
\subfigure[~The characteristics of the metric potential $g_{rr}$ when $v=0$ and $v\neq 0$]{\label{fig:1a}\includegraphics[scale=0.3]{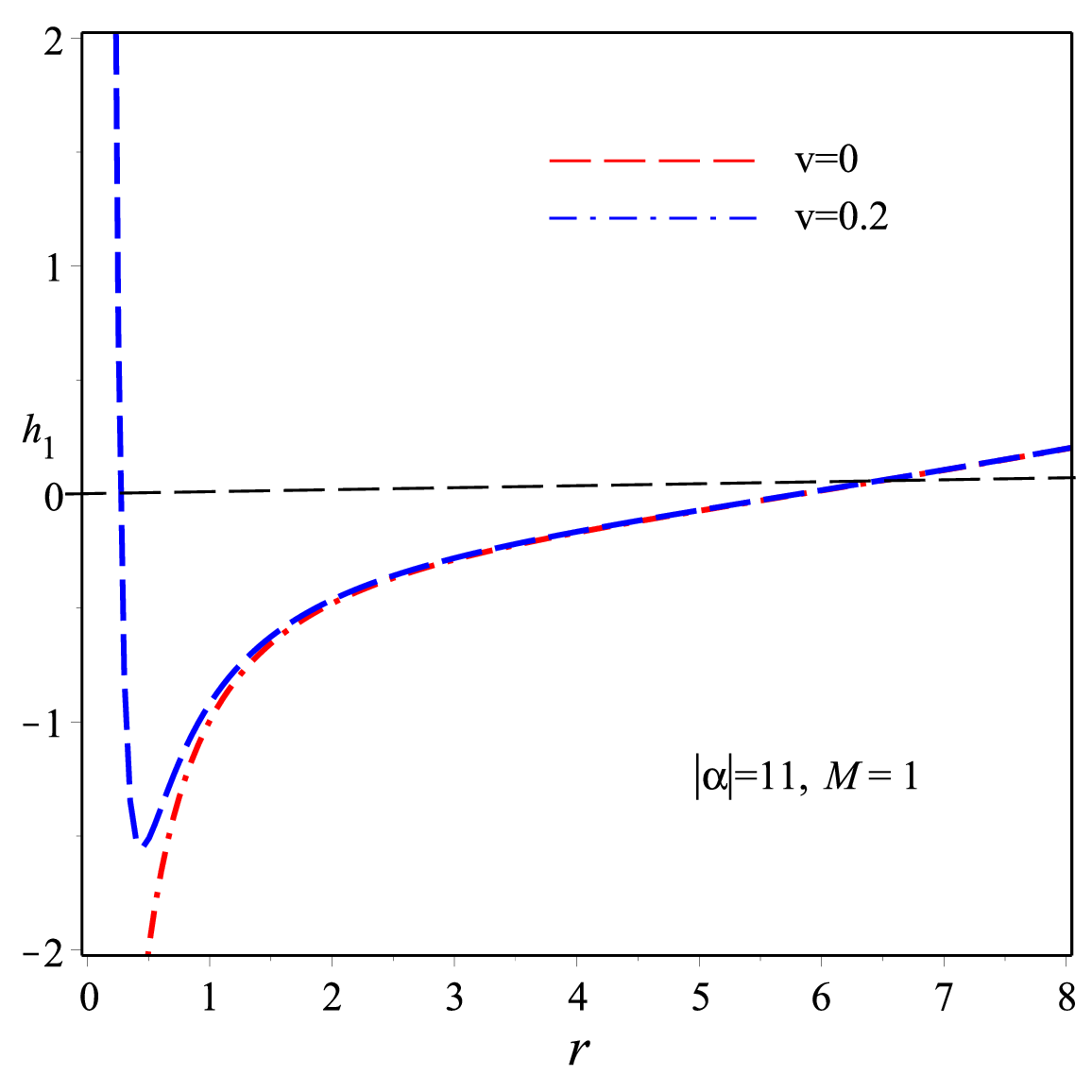}}\hspace{0.5cm}
\subfigure[~The locations of the horizons in the metric potential $g_{rr}$]{\label{fig:1b}\includegraphics[scale=0.3]{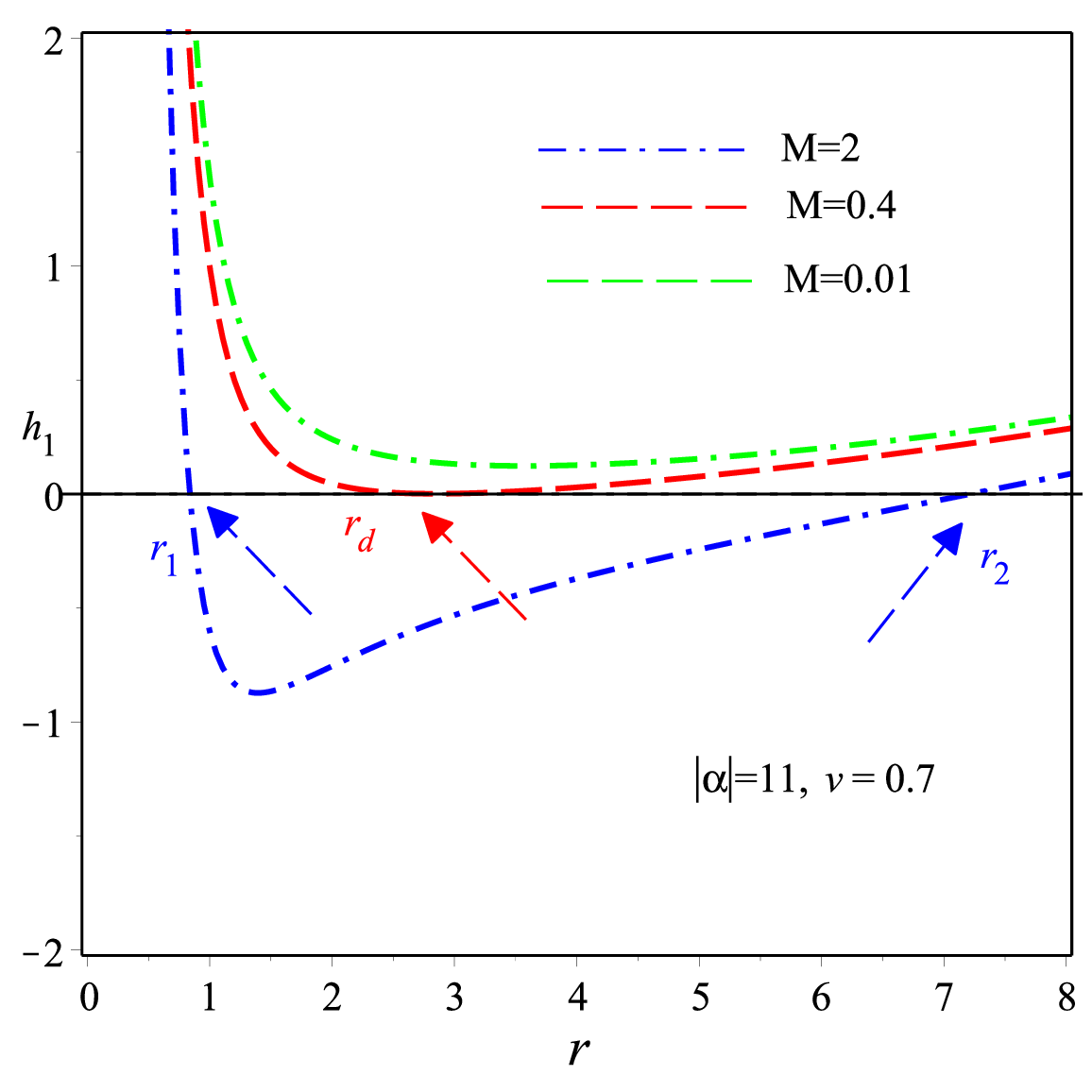}}
\subfigure[~The characteristics of the entropy]{\label{fig:1c}\includegraphics[scale=0.3]{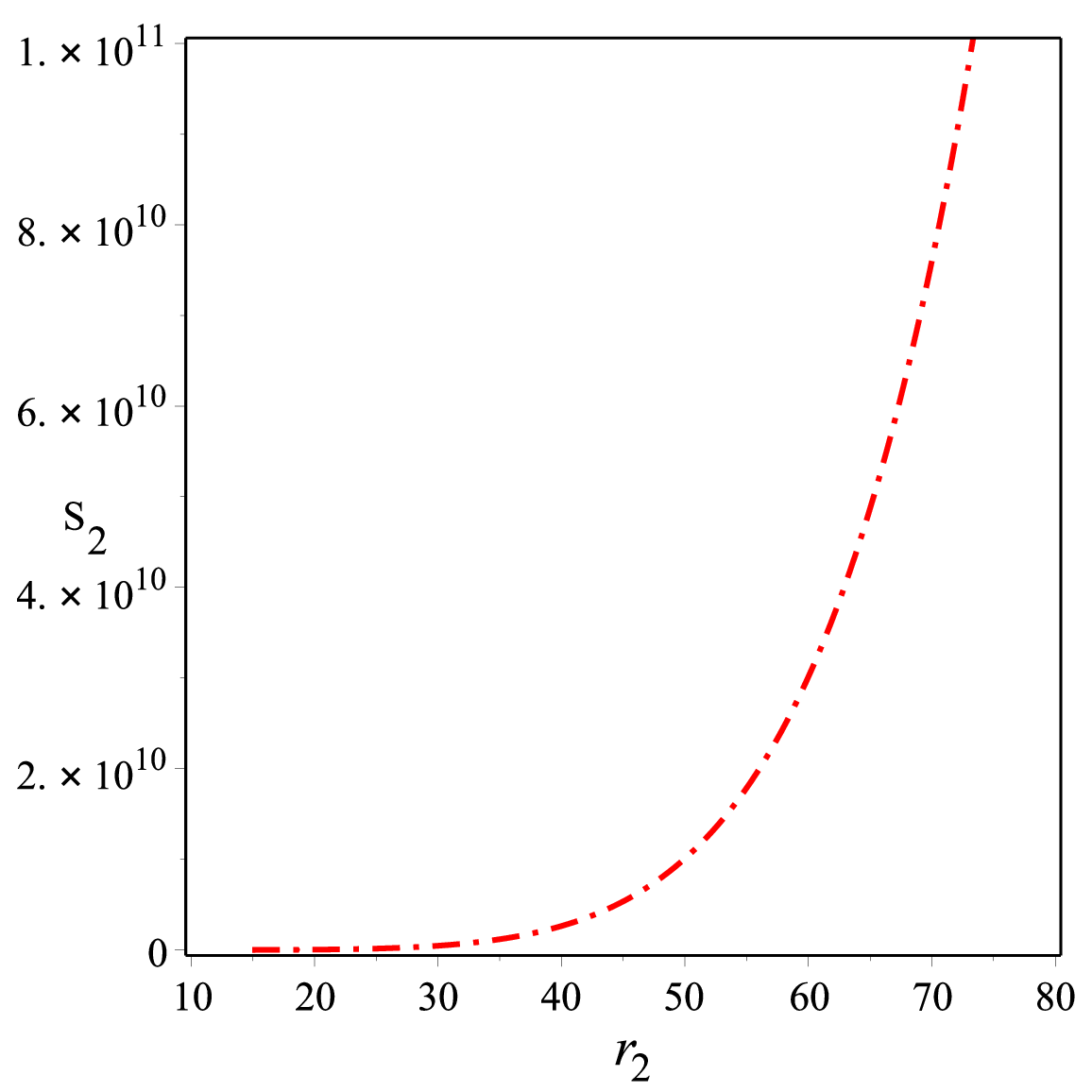}}
\subfigure[~The temperature according to Hawking]{\label{fig:temp}\includegraphics[scale=0.3]{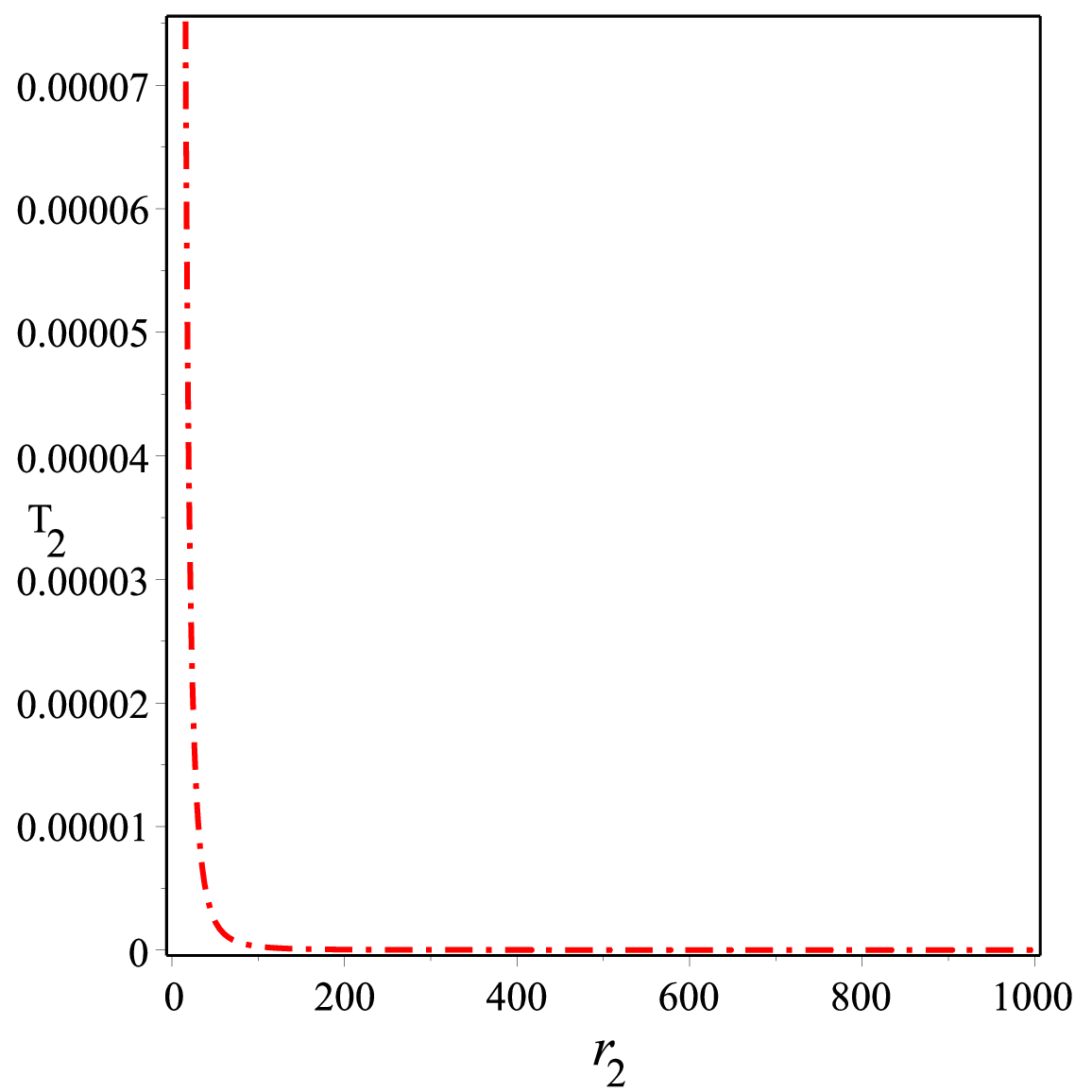}}\hspace{0.5cm}
\subfigure[~The characteristics of the Gibbs function]{\label{fig:gibb}\includegraphics[scale=0.3]{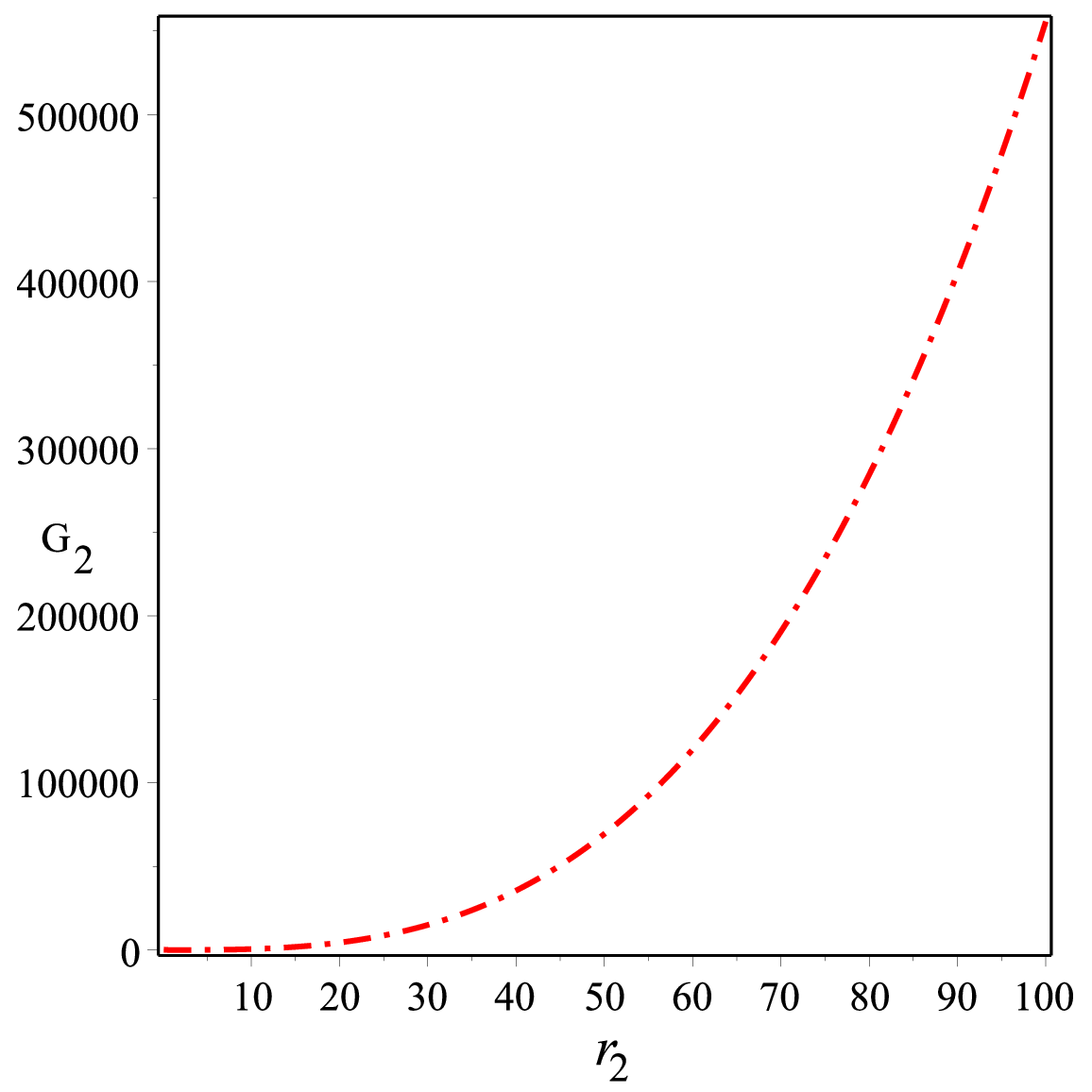}}\hspace{0.5cm}
\caption{\subref{fig:1a} The overall trends of the metric potential $g_{rr}$ are illustrated in Fig. \ref{Fig:Thr} for both $v=0$ and $v\neq 0$; \subref{fig:1b} presents the general tendencies of $g_{rr}$ as $r\to \infty$, revealing three distinct regions; \subref{fig:1c} showcases the behavior of the entropy; \subref{fig:temp} depicts the variations in the Hawking temperature; and \subref{fig:gibb} illustrates the pattern of the Gibbs function. { All the parameters characterized by the model take the values  $M=v=1$ and $\left|\alpha\right|=0.1$ for the calculations of thermodynamical quantities in Planck mass units}. }
\label{Fig:Thr}
\end{figure*}

{ In the context of the solution provided in Eq.~(\ref{solc}) and within the framework of four dimensions, i.e., $N=4$ which will be considered throughout the study of thermodynamics in section \ref{VI} and section \ref{VII}, the horizons consists of six roots whose algebraic equation can  be represented as:
\begin{align} \label{m33}
&r=\sqrt[6]{{z^6}+4v^3\sqrt{3\left|\alpha\right|^3}+18v^2z^2\left|\alpha\right|-12\left|\alpha\right|Mz^3}\,
\end{align}
where $z$ is the dependent variable. }
Additionally, from Eq.~(\ref{solc}), we can derive the following expression for mass:
\begin{align} \label{m44}
&M=\frac{r^6+4v^3\sqrt{3\left|\alpha\right|^3}+18\left|\alpha\right|v^2r^2}{12r^3}.
\end{align}
 Equation (\ref{m44}) illustrates that the horizon affects the black hole's total mass, $v$, and the dimensional constant $\left|\alpha\right|$. The relationship between  $h_1(r)$ and $r$ is depicted in Fig.\ref{Fig:Thr} \subref{fig:1a}, demonstrating the possible horizons of such solution for $v=0$ and $v\neq 0$. Moreover, Fig. \ref{Fig:Thr} \subref{fig:1b} shows the distinct areas of the horizons, where the blue curve represents the two horizons, the degenerate horizon is determined by equating $\frac{\partial M_h}{\partial r_h}$ to zero (depicted by the red curve), and the green curve represents the naked singularity zone when $M=0.3$.
Utilizing Eqs. (\ref{ent}) and (\ref{Inv}), we can determine the entropy associated with the described black hole solution of  Eq.~(\ref{solc}) as follows:
{ \begin{align} \label{ent1}
{S_2}=\frac {\pi {r_2}^{2}}{3}+\frac{8\pi v}{3}\,,
\end{align}
which suffers no divergence in the limit $v \to 0$ }.
The entropy trends are illustrated in Fig.\ref{Fig:Thr} \subref{fig:1c}, showcasing a well-behaved pattern of the entropy.

The Hawking temperature linked to the black hole solution of Eq.~(\ref{solc})   is calculated as:

\begin{align} \label{mt44}
&{T_2}=\frac{\left(\frac{{r_2}^{6}}{2\left|\alpha\right|}-4\,{v}^{3}\sqrt {3\left|\alpha\right|}-9\,{
v}^{2}{r_2}^{2}+3\,M{r_2}^{3} \right)}{2\pi {
r_2}^{5}\sqrt {
 \left( \frac{{r_2}^{2}}{{18\left|\alpha\right|}}+{\frac {{v}^{3}\sqrt {3\left|\alpha\right|}}{3{r_2
}^{4}}}+{\frac {3{v}^{2}}{2{r_2}^{2}}}-{\frac {M}{r_2}} \right) ^{2}
 \left( 1+{\frac {v\sqrt {3}\sqrt {\alpha}}{{r_2}^{2}}} \right) ^{2}}}\,.
\end{align}
Here ${T_1}$ represents the temperature of  Hawking at the event horizon. The temperature is depicted in Fig. \subref{fig:temp}, revealing that it is always positive.

The grand canonical ensemble's free energy, which is  known as Gibbs free energy, is described as \cite{Zheng:2018fyn, Kim:2012cma}:
\begin{equation} \label{enr}
G(r_2)=M(r_2)-T(r_2)S(r_2),
\end{equation}
where $T(r_2)$, $S(r_2)$, and $E(r_2)$ represent the temperature, entropy, and quasilocal energy at the event horizon, respectively.  By substituting Eqs. (\ref{ent}), (\ref{m33}), (\ref{ent1}), and (\ref{m44}) into (\ref{enr}), we obtain:
\begin{align} \label{m77}
&G(r_2)=\left\{702{v}^{6}\sqrt {3\left|\alpha\right|^5}{r_2}^{4}-48{r_2}^{6}{v}^{4}\sqrt {3\left|\alpha\right|^5}-108{r_2}^{6}{v}^{2}\sqrt { 3\left|\alpha\right|^5}+36{r_2}^{7}M\sqrt {3\left|\alpha\right|^5}+55{r_2}^{8}{v}^{4}\sqrt {3\left|\alpha\right|^3}-12{r_2}^{9}M{v}^{2}\sqrt {3\left|\alpha\right|^3}\right.\nonumber\\
&\left.-396\,{v}^{4} \sqrt {3\left|\alpha\right|^5}M{r_2}^{5}+v{r_2}^{14}+45\,{r_2}^{10}{v}^{3}\left|\alpha\right|+ 516\,{v}^{5}{\left|\alpha\right|}^{2}{r_2}^{6}+6\,\left|\alpha\right|\,{r_2}^{12}v-108\,{\left|\alpha\right|}^{2} {r_2}^{8}{v}^{3}+720\,{v}^{7}{\left|\alpha\right|}^{3}{r_2}^{2}-144\,{\left|\alpha\right|}^{3}{r_2}^{4 }{v}^{3}\right.\nonumber\\
&\left.+72\,{v}^{8}\sqrt {3{\left|\alpha\right|}^7}-12\,v{r_2}^{11}M\left|\alpha\right|-324\, {v}^{3}{\left|\alpha\right|}^{2}{r_2}^{7}M+36\,{\left|\alpha\right|}^{2}{r_2}^{9}Mv-216\,{v}^{5}{ \left|\alpha\right|}^{3}M{r_2}^{3}+6\,{r_2}^{10}\sqrt {3{\left|\alpha\right|}^3}+{r_2}^{12}{v}^{2} \sqrt {3\left|\alpha\right|}\right\}\nonumber\\
&\left\{18\left|\alpha\right|\,{r_2}^{3}v \left( {r_2}^{6}+4\,{v}^{3} \sqrt {3\left|\alpha\right|^3}+18\,{v}^{2}\left|\alpha\right|\,{r_2}^{2}-12\,M\left|\alpha\right|\,{r_2}^{3 } \right)  \left( {r_2}^{2}+v\sqrt {3\left|\alpha\right|} \right) \right\}^{-1}\,.
\end{align}

The trends of the Gibbs energy for the black hole (\ref{solc}) are depicted in Fig. \ref{Fig:Thr}\subref{fig:gibb} for specific values of the model parameters. Fig. \ref{Fig:Thr}\subref{fig:gibb} illustrates that the Gibbs free energy is consistently positive, indicating greater global stability \cite{Nashed:2021pah}.
\section{Solutions with multiple horizons}\label{VII}
The most basic black hole solution is characterized by the Ads/dS-Schwarzschild metric, where the description of the metric coefficient $g_{00}$ is:
\begin{equation}
\label{hor111}
f= \Lambda r^2-\frac{2M}{r}=f_1(r)(r-r_1)(r-r_2)(r-r_3)\,.
\end{equation}
Here, $f_1(r)=\frac{1}{r}$ and $r_i, i=1,2,3$ represent the horizons. Equation~(\ref{hor111}) possesses a single real horizon at $r=r_1=\sqrt[3]{2M\Lambda}$, while the other two horizons, $r_2$ and $r_3$, are imaginary.
Equation~(\ref{hor111}) can be derived from Eq.~(\ref{solc}) when $v=0$. When $v\neq 0$, we derive the solution of Eq. (\ref{solc}) in the case of four dimensions in the following manner:
{ \begin{equation}
\label{hor222}
h=\frac{r^2}{12\alpha}-\frac{M}{r}+\frac{3v^2}{2r^{2}}+\frac{\sqrt{3\left|\alpha\right|}v^3}{3
r^4}=h_2(r)(r-r_1)(r-r_2)(r-r_3)(r-r_4)(r-r_5)(r-r_6)\,.
\end{equation}}
 We identify the impact of the quadratic form of non-metricity, generating six roots from which at least three can be extracted where, $h_2(r)=\frac{1}{r^4}$.
Now we are going to show that, it is possible to produce at least three real roots of Eq.~(\ref{hor222}) in the following manner:
{\begin{equation}
\label{hor333}
h=\frac{r^2}{12\alpha}-\frac{M}{r}+\frac{3v^2}{2r^{2}}+\frac{\sqrt{3\left|\alpha\right|}v^3}{3
r^4}=h_2(r)(r-r_1)(r-r_2)(r-r_3)\,.
\end{equation}}
In this study, $r_1$ and $r_2$ stand for the cosmological constant\footnote{ The negative cosmological constant is crucial in the AdS/CFT correspondence, which establishes a duality between a quantum gravity theory in anti-de Sitter space and a conformal field theory in one lower dimension. This duality offers a distinctive approach for studying strongly coupled field theories by leveraging classical gravity \cite{Giambo:2023zmy}.} and event horizons, respectively, while $r_3$ is the radius horizon obtained from the contribution of the quadratic form of non-metricity. Deriving the explicit forms of the three horizons for Eq.~(\ref{solc}) poses a challenge.
Hence, we will numerically and graphically solve Eq.~(\ref{solc}) in cases where $v<0$.
We graph Eq.~(\ref{solc}) regarding certain values of mass, charge, and parameter  $\alpha$  related to the quadratic form of non-metricity.

\begin{figure}[ht]
\centering
\subfigure[~The BH's multi-horizon]{\label{fig:hori}
\includegraphics[scale=0.35]{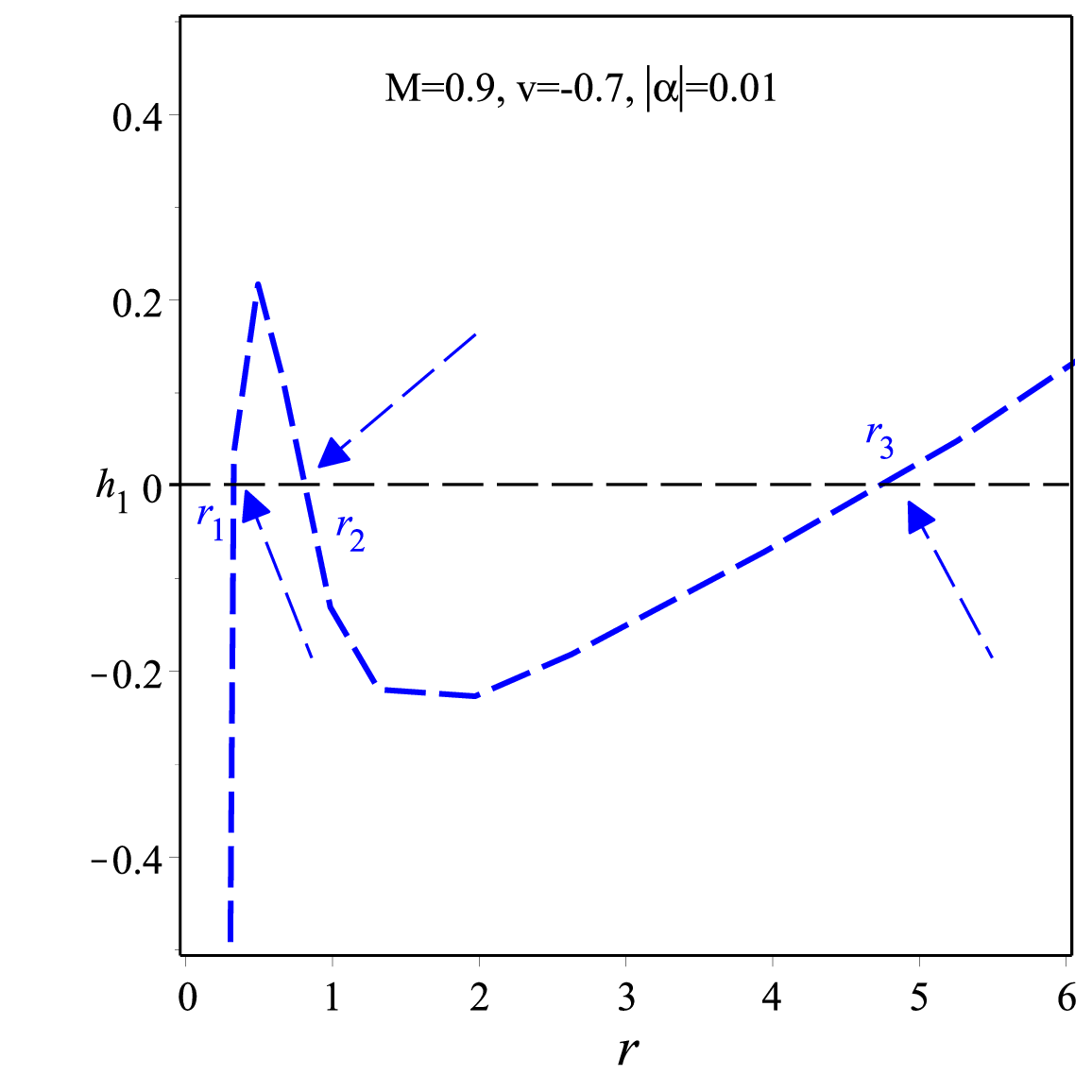}}\hspace{0.2cm}
\caption{ { Multi-horizon plot of Eq.~(\ref{solc}) against coordinate $r$ in which all the parameters model are in Planck mass units}.}
\label{Fig:hor}
\end{figure}
In the case of a black hole with one or two horizons, the curvature scalar is recognized to be zero, as observed in the Schwarzschild and Reissner-Nordstr\"om solutions. However, for a black hole with more than two horizons, the curvature scalar becomes singular at the center. At $r=0$, the Kretschmann scalar is singular for any black hole with more than one horizon. This is explained in detail by Eq.~(\ref{Inv}).

\begin{figure*}
\centering
\subfigure[~The Hawking temperature associated with the horizon of the BH described in Eq.~(\ref{solc})]{\label{fig:3a}\includegraphics[scale=0.3]{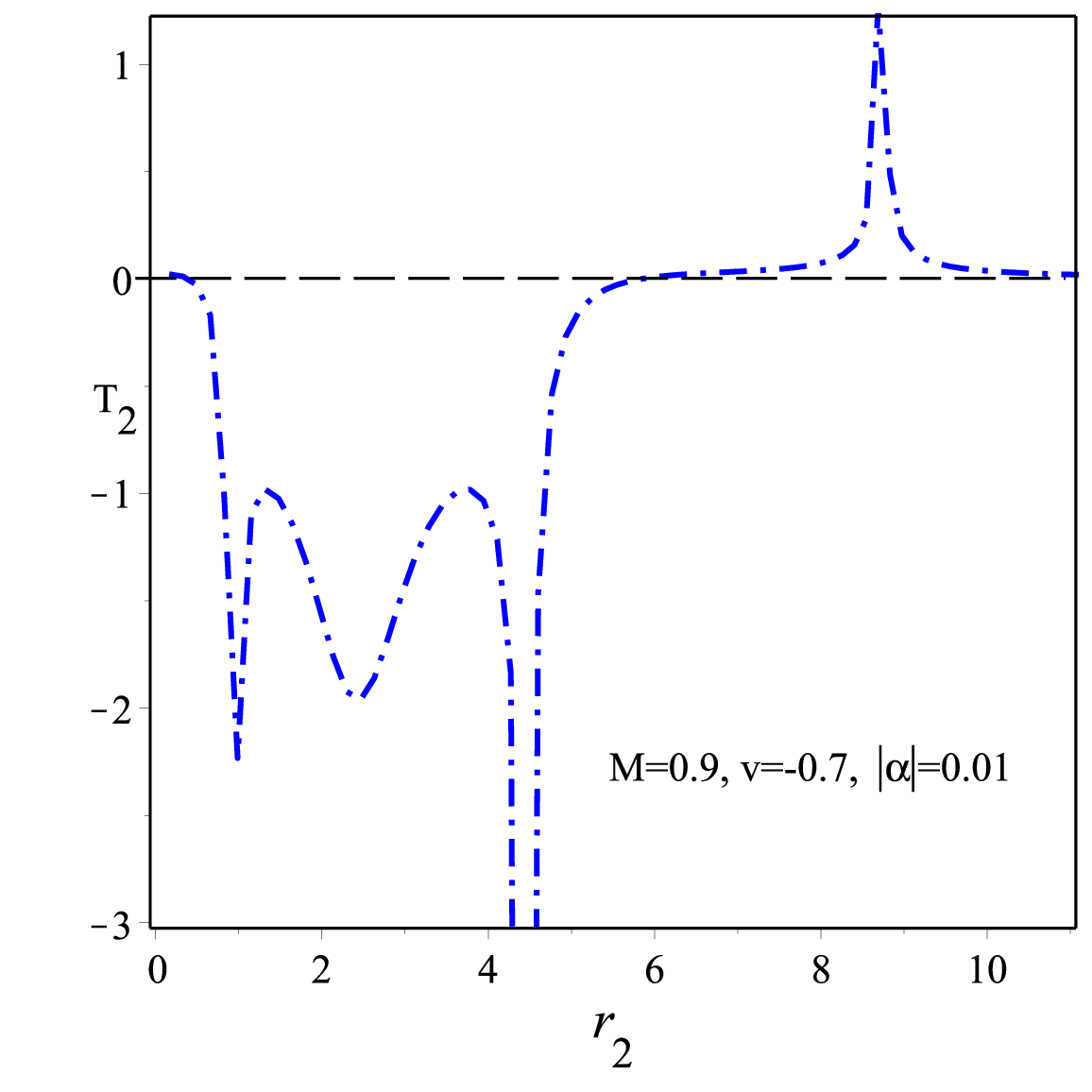}}\hspace{0.5cm}
\subfigure[~The the Gibbs function of solution described in Eq.~(\ref{solc})]{\label{fig:3b}\includegraphics[scale=0.3]{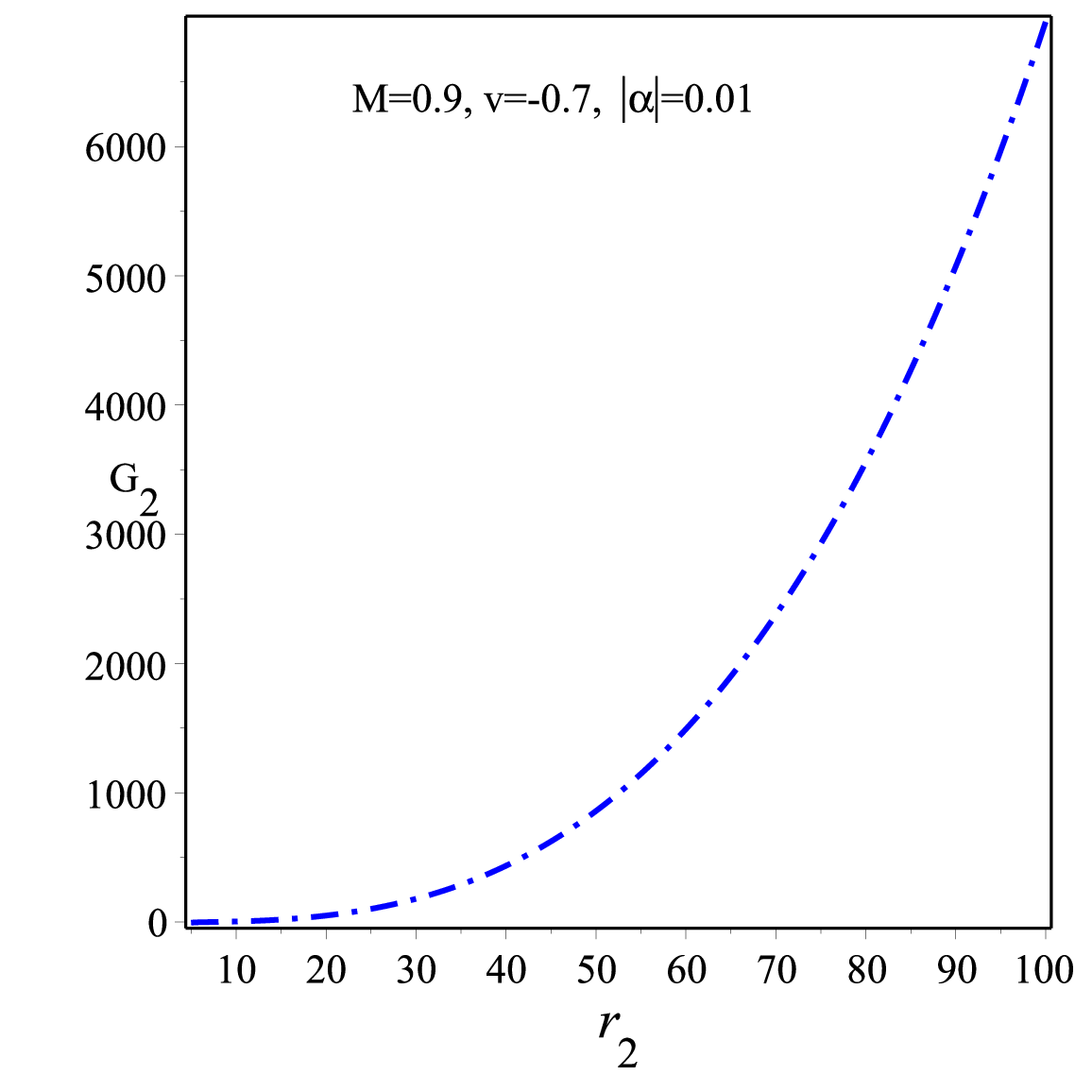}}\hspace{0.5cm}
\caption{Thermodynamic features of the BH solutions (\ref{solc}) illustrated in diagrams for parameter $v$ taking negative values:~\subref{fig:3a}
the characteristic pattern of the black hole's horizon Hawking temperature (\ref{solc}) and~\subref{fig:3b} { the Gibbs function of the black hole's horizon (\ref{solc}). All the parameters of the model are in Planck mass units.}}
\label{Fig:3}
\end{figure*}

Let's now examine the thermodynamics of the multi-horizons stated earlier.
Because of the negative value of $v$ in the case of multi-horizons, we obtain for BH (\ref{solc}) that these quantities behave differently thermodynamically.
We have stable BHs for a multi-horizon spacetime, as demonstrated by all the charts in Fig.~\ref{Fig:3}.  The Hawking temperature in the multi-horizon situation is negative up to $r<r_d$ and then positive as $r>r_d$, as can be shown in Fig.~\ref{Fig:3} \subref{fig:3a}.
\section{Conclusions and discussion}\label{VIII}

The study of charged cylindrical black holes within the framework of $f(\mathbb{Q})$ gravitational theory has provided significant insights into these unique astronomical objects. This research presents a comprehensive investigation into how modifications in $f(\mathbb{Q})$ influence the properties and behavior of charged cylindrical black holes, offering a valuable understanding of the interaction between gravity and this modified theory.


We present a novel charged solution within the framework of Maxwell-$f(\mathbb{Q})$ gravitational theory for dimensions $N\geq 4$, with the coincident gauge condition imposed \cite{Heisenberg:2023lru}\footnote{The coincident gauge is adopted here because it simplifies the geometry by aligning the coordinates directly with physical observables, reducing the complexity of the connection. This results in equations that are more tractable and suited to the system's symmetry. However, if the coincident gauge introduces complications, such as non-intuitive behavior or singularities, an alternative gauge, such as the Lorenz gauge (or harmonic gauge in gravitational contexts), might be preferable to maintain symmetry without these issues. Alternatively, focusing on gauge-invariant quantities could avoid dependence on specific gauge choices.}. For $f(\mathbb{Q})=\mathbb{Q}+\frac{1}{2}\alpha \mathbb{Q}^2$, where $\alpha <0$, we derive a specific solution. This solution exhibits intriguing features, including both quadrupole and monopole terms. This black hole approaches Anti-de Sitter (AdS) space.

We present a novel charged solution within the framework of Maxwell-$f(\mathbb{Q})$ gravitational theory for dimensions $N \geq 4$, with the coincident gauge condition imposed \cite{Heisenberg:2023lru}\footnote{The coincident gauge is adopted here because it simplifies the geometry by aligning the coordinates directly with physical observables, thereby reducing the complexity of the connection. This results in equations that are more tractable and better suited to the system's symmetry. However, if the coincident gauge introduces complications, such as non-intuitive behavior or singularities, an alternative gauge such as the Lorenz gauge (or harmonic gauge in gravitational contexts) might be preferable to maintain symmetry without these issues. Alternatively, focusing on gauge-invariant quantities could avoid dependence on specific gauge choices.}.  For $f(\mathbb{Q}) = \mathbb{Q} + \frac{1}{2}\alpha \mathbb{Q}^2$, where $\alpha < 0$, we derive a specific solution that exhibits intriguing features, including both quadrupole and monopole terms. This black hole solution asymptotically approaches Anti-de Sitter (AdS) space.
From a physical standpoint, these entities can contribute to the ongoing debate about which gravity model is the most reliable, whether based on curvature or non-metricity. As noted in the literature, since GR is largely equivalent in both frameworks, the discussion should focus on the distinctions between $f(\mathbb{Q})$ and $f(\mathbb{R})$ constructions, as they fundamentally differ. Even basic phenomena, such as gravitational waves, exhibit significant differences between the $f(\mathbb{Q})$ and $f(\mathbb{R})$ formulations, as shown in \cite{Soudi:2018dhv}. In resolving this debate, a deeper understanding of black hole properties could prove invaluable.

{D'Agostino et al. \cite{DAgostino:2024ymo} investigated the impact of geometric and topological corrections to Einstein's gravity on black hole thermodynamics. They first examined a logarithmic correction to the Hilbert-Einstein action, which results from a small power-law perturbation of the Ricci scalar, consistent with current experimental constraints on general relativity. Additionally, they explored a topological correction introduced by adding the squared Gauss-Bonnet term to the gravitational action. In both scenarios, they assumed that these correction terms modify the spherically symmetric spacetime lapse function. In the first scenario, they showed that the Minkowski limit cannot be recovered, while in the second case, the modified solution decays more rapidly than the Schwarzschild solution.

In our study, we depart from spherical symmetry and instead adopt cylindrical symmetry. Our analysis is broadly applicable to the $f(\mathbb{T})$ and $f(\mathbb{Q})$ theories. In the quadratic case, where
$h=h_1$, the correction behaves similarly to a cosmological term. For the charged scenario, our results differ from the topological correction considered by D'Agostino et al. \cite{DAgostino:2024ymo}, which involves the square of the Gauss-Bonnet term. However, when the linear form of the Gauss-Bonnet term is used, the correction behaves like a topological term. Unlike their study, our model includes only the Maxwell field, which is not topological. This distinction ensures that, in the $f(\mathbb{Q})$ theory, the Reissner-Nordstr\"om solution cannot be recovered if the correction term vanishes.

In the case of anti-de-Sitter spacetime, it was demonstrated by examining the $t,t$-component \cite{Giambo:2023zmy} and applying the condition of isotropic pressure, which ensures that the external dark energy responsible for the temperature field aligns with key features observed in current measurements. They \cite{Giambo:2023zmy} showed that the nature of their solutions, whether regular or singular, depends on the sign of the cosmological constant: negative for regular solutions and positive for singular ones. By varying the cosmological constant, they \cite{Giambo:2023zmy} were able to derive both anti-de Sitter and de Sitter-like behaviors. The applicability of this approach to the context of the present study will be explored in future work.}
\section*{Acknowledgements}
I would like to thank S. Nojiri for the useful discussion during the preparation of this manuscript and Kobayashi-Moskawa Institute for hosting me while doing this study.

\end{document}